\documentclass[onecolumn,prd,nofootinbib,showpacs]{revtex4}

\usepackage{textcomp}
\usepackage{mathtools, ulem, graphicx}
\usepackage[export]{adjustbox}
\usepackage{sidecap}
\usepackage{verbatim}
\newcommand{\e}{\begin{equation*}\begin{aligned}}
\newcommand{\ee}{\end{aligned}\end{equation*}}

\newcommand{\en}{\begin{equation}\begin{aligned}}
\newcommand{\een}{\end{aligned} \end{equation}}

\newcommand{\p}{\partial}

\newcommand{\f}[2]{\frac{#1}{#2}}

\newcommand{\ra}{\rangle}
\newcommand{\la}{\langle}

\newcommand{\da}{\dagger}
\newcommand{\ma}{\mathcal}

\newcommand{\Q}{\left}
\newcommand{\W}{\right}

\newcommand{\pma}{\begin{pmatrix}}
\newcommand{\epma}{\end{pmatrix}}

\newcommand{\na}{\nabla}
\newcommand{\de}{\delta}

\newcommand{\be}{\beta}
\newcommand{\lam}{\lambda}

\begin{document}

\title{Virial Theorem for Non-relativistic Quantum Fields in D Spatial Dimensions}
\author{Chris L. Lin}
\affiliation{Department of Physics, University of Houston, Houston, TX 77204-5005}

\author{Carlos R. Ord\'{o}\~{n}ez}
\affiliation{Department of Physics, University of Houston, Houston, TX 77204-5005}
\affiliation{Department of Science and Technology, Technological University of Panama, Campus Victor Levi, Panama City, Panama.}

\date{\today}
\email{cllin@uh.edu}
\email{cordonez@central.uh.edu}
%\begin{abstract}
%We give a brief introduction to the use of \LaTeX\ in the context of REVTeX~4.1.
%\end{abstract}


\begin{abstract}
The virial theorem for non-relativistic complex fields in $D$ spatial
dimensions and with arbitrary many-body potential is derived, using
path-integral methods and scaling arguments recently developed to
analyze quantum anomalies in low-dimensional systems. The potential
appearance of a Jacobian  $J$  due to a change of variables in the
path-integral expression for the partition function of the system is
pointed out, although in order to make contact with the literature most
of the analysis deals with the $J=1$ case.  The virial theorem is recast
into a form that displays the effect of microscopic scales on the
thermodynamics of the system. From the point of view of this paper the
case usually considered, $J=1$, is not natural, and the generalization to
the case $J\neq1$ is  briefly presented.
\end{abstract}

\maketitle

\section{Introduction}

The virial theorem has been proven using a variety of methods. Recently, a path-integral derivation of the virial theorem has been developed in the context of quantum anomalies in non-relativistic 2D systems, or more generally, systems with $SO(2,1)$ classical symmetry \cite{Ord}.  The path integral is most useful in isolating the anomaly contribution to the equation of state so obtained. This method is in fact quite general, and applicable for non-relativistic systems with an arbitrary 2-body potential $V(\vec{x}_1,\vec{x}_2)$ in $D$ spatial dimensions, even when there are no quantum anomalies present. We present such derivation in this note, extending the original derivation using also diagrammatic analysis, and recasting the virial theorem into a general equation that relates macroscopic thermodynamics variables to the microscopic physics.  As it will be shown, there is generically a Jacobian term $J$ that may contribute to the virial theorem, regardless of the existence of a classical scaling symmetry. We will mainly concern ourselves here with the case $J=1$ (which we term  ``non-anomalous''). Comments and conclusions end the note. 

\begin{comment}The virial theorem has been proven with many different methods []. We present here a path-integral proof in the grand canonical ensemble. The merit of the path-integral approach is the ability for it to catch anomalies, something standard proofs completely miss [true???].  Indeed, in previous work we focused exclusively on potentials that are scale invariant [], and the calculation of anomalies [] via path integrals for these scale-invariant systems. However, in this paper we focus on the non-relativistic, non-anomalous virial theorem in D dimensions for general 2-body potential. We will show that even if anomalies are not taken into account, the path-integral provides a relatively simple proof of the virial theorem for arbitrary D-dimensional, 2-body potential. Furthermore, we recast the virial theorem into a general thermodynamic equation that relates macroscopic thermodynamic variables to the microscopic physics.
\end{comment}

\section{Virial Theorem}

The work in \cite{Ord} was based partly on the work by Toyoda et al. \cite{t1,t2,t3}. They postulated that spatial scalings\footnote{Toyoda et al. introduced an auxiliary external potential that has the effect of confining the system to a volume $V$, and then, through a series of infinitesimal scalings and algebraic arguments derived what amounts to the equation of state, which they referred to as virial theorem.  Unlike them, we're not using an external potential but simply consider a system with a large volume $V$(so all the typical large-volume thermodynamical considerations apply), but like them, we're also calling virial theorem the equation of state that will be derived in this paper. }   

\en
\vec{x}\,'&=\lam \vec{x}\, , \\
\psi'(t,\vec{x}\,')&=\lambda^{-D/2}\psi(t,\vec{x}\,),
\een

leave the particle number density invariant:

\en
d^D\vec{x} \, \psi^\da(t,\vec{x}\,)\psi(t,\vec{x}\,)=d^D\vec{x}\,' \,\psi'^\da(t,\vec{x}\,'\,)\psi'(t,\vec{x}\,'\,).
\een

Let us consider a non-relativistic system whose microscopic physics is represented by a generic 2-body interaction\footnote{In this paper we set $\hbar=m=1$.}

\en
\ma L=\psi^* \Q( i\p_t+\f{\nabla^2}{2}\W)\psi-\f{1}{2} \int d^D \vec{y} \,\psi^*(t,\vec{x}) \psi(t,\vec{x})V(\vec{x}-\vec{y}\,)\psi^*(t,\vec{y}) \psi(t,\vec{y}).
\een

Giving our system a macroscopic volume $V$, temperature $\be^{-1}$, and chemical potential $\mu$, and going into imaginary time gives for the partition function:

\en
Z[V,\be,\mu]=\int [d\psi^*][d\psi] e^{-\int_0^\beta d\tau \int_V d^D\vec{x}\, \Q[ \psi^* \Q( \p_\tau-\f{\nabla^2}{2}-\mu\W)\psi+\f{1}{2} \int d^D \vec{y} \,\psi^*(\tau,\vec{x}) \psi(\tau,\vec{x})V(\vec{x}-\vec{y}\,)\psi^*(\tau,\vec{y}) \psi(\tau,\vec{y})\W]}.
\een

Now consider a new system with the same temperature and chemical potential, but at volume $V'=\lambda^D V$:

\en
Z[\lambda^D V,\be,\mu]=\int [d\psi'^*][d\psi'] e^{-\int_0^\beta d\tau \int_{\lambda^D V}d^D\vec{x}' \Q[ \psi'^* \Q( \p_\tau-\f{\nabla'^2}{2}-\mu\W)\psi'+\f{1}{2} \int d^D \vec{y}\,' \,\psi'^*(\tau,\vec{x}\,') \psi(\tau,\vec{x}\,')V(\vec{x}\,'-\vec{y}\,'\,)\psi'^*(\tau,\vec{y}\,') \psi'(\tau,\vec{y}\,')\W]}.
\een

%It should be emphasized that $Z^\lambda[\lambda^D V,\be^{-1},\mu] \neq Z^1[\lambda^D V,\be^{-1},\mu]$ because the microscopic physics has changed (this is due to the lack of symmetry of the action under Eq. (3)). 

Substituting Eq. (1) into Eq. (5) gives:

\en
Z[\lambda^D V,\be,\mu]=\int [d\psi^*][d\psi] J e^{-\int_0^\beta d\tau \int_{V}d^D\vec{x} \Q[ \psi^* \Q( \p_\tau-\f{1}{\lam^2}\f{\nabla^2}{2}-\mu\W)\psi+\f{1}{2} \int d^D \vec{y} \,\psi^*(\tau,\vec{x}\,) \psi(\tau,\vec{x}\,)V(\lambda \Q(\vec{x}\,-\vec{y}\W)\,)\psi^*(\tau,\vec{y}\,) \psi(\tau,\vec{y}\,)\W]},
\een

where $J$ is the Jacobian for the transformation $(\psi'^*,\psi')\rightarrow(\psi^*,\psi)$. As mentioned above, our emphasis will be in the non-anomalous case, and henceforth we assume $J=1$ (see however comments and conclusions).  Then $Z[\lambda^D V,\be,\mu]\equiv Z^\lam[V,\be,\mu]$, where the superscript $\lam$ represents a microscopic system whose kinetic energy has a factor $\f{1}{\lambda^2}$ and whose potential is $V(\lambda \Q(\vec{x}\,-\vec{y}\W)\,)$. Note that $Z^{\lam=1}[V,\be,\mu]=Z[V,\be,\mu]$.\\

The pressures corresponding to $Z[\lambda^D V,\be,\mu]$ and $Z[V,\be,\mu]$ are equal, since the intensive variables $\mu$ and $\be^{-1}$ are the same, and they correspond to the same microscopic system. %(the action in Eq. (5) is the same action as in Eq. (4) since the fields are dummy indices). 
The argument we just made for the pressures being the same is valid in the thermodynamic limit, based on the principle that two intensive variables determine the third via an equation of state e.g., $P=\rho T$ for an ideal gas. However, in the next section we will also provide a diagrammatical proof that the two pressures are the same.   \\

For now assume the pressures are equal. Then using $Z=e^{\beta PV}$, we get:

\en
e^{\be PV'}-e^{\be PV}&=Z[\lambda^D V,\be,\mu]-Z[V,\be,\mu],\\
\text{or} \quad e^{\be P\lam^D V}-e^{\be PV}&=Z^\lam[V,\be,\mu]-Z[V,\be,\mu].
\een

Following \cite{Ord}, we set $\lam=1+\eta$ for infinitesimal $\eta$:

\en
e^{\be P V}D\eta \beta PV&=Z^{\lam=1}[V,\be,\mu]+\p_\lam Z^\lam[V,\be,\mu] \Big |_{\lam=1}\eta-Z[V,\be,\mu]\\
&=\p_\lam Z^\lam[V,\be,\mu] \Big |_{\lam=1}\eta\\
&=Z[V,\be,\mu]\Q\la \int_0^\beta d\tau \int_{V}d^Dx\,\Q( -\psi^\da\nabla^2\psi -\f{1}{2}\int d^D \vec{y}\, \rho(\tau,\vec{y}\,) \Q[(\vec{x}-\vec{y}\,)\cdot\na_{\vec{x}} V(\vec{x}-\vec{y}\,) \W]\rho(\tau,\vec{x}\,)\W)\W\ra\eta,
\een

where we've defined $\rho(\tau,\vec{x}\,)\equiv \psi^\da(\tau,\vec{x}) \psi(\tau,\vec{x})$. Cancelling the partition functions on both sides, noting that thermal expectation values for the fields at the same $\tau$ are independent of $\tau$ so that the $\tau$ integral pulls out a $\beta$, and denoting the kinetic energy as $KE$:

\en
DPV=2KE-\Q\la \f{1}{2}\int d^D \vec{x}\,\int d^D \vec{y}\,\rho(\tau,\vec{y}\,)  \Q[(\vec{x}-\vec{y}\,)\cdot \na_{\vec{x}} V(\vec{x}-\vec{y}\,)  \W] \rho(\tau,\vec{x}\,)  \W\ra,
\een

which is the virial theorem in $D$ dimensions (Eqs. (3.30) and (2.6) in \cite{t2} and \cite{t3} respectively).

\section{N-body}

It is clear that this method can be generalized to the n-body case. Since by Eq. (2) the scaling transformation preserves $\int d^D \vec{x} \,\psi^\da (\tau,x) \psi(\tau,x)$ ($\equiv \int d^D \vec{x} \, \rho(\tau,\vec{x}\,)$), an n-body term transforms as

\en
\f{1}{n!}\int \Q(\prod \limits_{i}^{n} d^D \vec{x}_i \,  \rho(\tau,\vec{x}_i\,)\W) V(\vec{x}_1, ...,\vec{x}_n ) \rightarrow \f{1}{n!}\int \Q(\prod \limits_{i}^{n} d^D \vec{x}_i  \,\rho(\tau,\vec{x}_i\,) \W) V(\vec{x}_1', ...,\vec{x}_n' ).
\een

Setting $V(\vec{x}_1, ...,\vec{x}_n)=\tilde{V}(\vec{z}_{\text{COM}}, \vec{z}_2,...,\vec{z}_n)$ where $\vec{z}_i\equiv \vec{x}_i-\vec{x}_1$ and $\vec{z}_{\text{COM}}$ is the center of mass of the $\vec{x}_i$'s gives 

\en
DPV=2KE-&\Q\la \f{1}{n!} \int \Q(\prod \limits_{i}^{n} d^D \vec{x}_i \,\rho(\tau,\vec{x}_i\,)\W) \Q[\vec{z}_{\text{COM}}\,\cdot \na_{\vec{z}_{\text{COM}}} \tilde{V}(\vec{z}_{\text{COM}}, \vec{z}_2,...,\vec{z}_n)  \W]   \W\ra \\ -&\Q\la \f{1}{n!} \int \Q(\prod \limits_{i}^{n} d^D \vec{x}_i \,\rho(\tau,\vec{x}_i\,)\W) \Q[\sum\limits_{i=2}^{n}\vec{z}_i\,\cdot \na_{\vec{z}_i} \tilde{V}(\vec{z}_{\text{COM}}, \vec{z}_2,...,\vec{z}_n)  \W]   \W\ra.
\een

For translationally-invariant systems, we can ignore the potential term in the 1st line.

\section{Diagrammatic Proof of \textit{P}=\textit{P'}}

To prove diagramatically that the pressure $P'$ corresponding to $Z[\lambda^D V,\be,\mu]$ is equal to the pressure $P$ corresponding to $Z[V,\be,\mu]$, it suffices to show that $\Omega[\lambda^D V,\be,\mu]=\lambda^D \Omega[V,\be,\mu]$, where $\Omega$ is the grand potential. By the cluster expansion, $\Omega$ is given by the sum of connected vacuum graphs \cite{Wein}. Using the Feynman rules, $\Omega[V,\be,\mu] \propto \delta^D(0) \ma M(\be,\mu)$, where $\delta^D(0)$ expresses conservation of momentum of the vacuum and $\ma M(\be,\mu)$ is the Feynman amplitude\footnote{$\ma M$ is the T-matrix, and $\delta^D(0)=\int  \f{d^Dx}{(2\pi)^D} \,e^{-i0*x} \propto V$.} which is independent of $V$, since $\ma M$ contains expressions like $\f{\Delta n_1...\Delta n_D}{V} f\Q(\f{2\pi n_i}{L}\W)$ which in the continuum limit $\rightarrow \f{d^Dk}{(2\pi)^D} f\Q(k_i\W)$\footnote{For finite volume, momenta are discrete and summed over: $k_i=\f{2\pi n_i}{L}$. $\Delta n_1...\Delta n_D$ is a box of unit volume surrounding the discrete lattice point $n_i$. In the limit of large $L$, $f\Q(\f{2\pi n_i}{L}\W)$ is assumed not to vary much, so any point within $\Delta n_1...\Delta n_D$ not on the lattice would still contribute the same value of $f\Q(\f{2\pi n_i}{L}\W)$. Then $\sum\limits_{n_i}\f{1}{V} f\Q(\f{2\pi n_i}{L}\W)=\sum\limits_{n_i}\f{\Delta n_1...\Delta n_D}{V} f\Q(\f{2\pi n_i}{L}\W) \rightarrow \int \f{dn_1...dn_D}{V} f\Q(\f{2\pi n_i}{L}\W) = \int \f{d^Dk}{(2\pi)^D} f\Q(k_i\W)$.}. 

Taking $\delta^D(0) \propto V$, it's clear that $\Omega[V,\be,\mu] \propto V \ma M(\be,\mu)$, so $\Omega[\lambda^D V,\be,\mu]=\lambda^D \Omega[V,\be,\mu]$ in the continuum limit.  \\

Alternatively since $Z[\lambda^D V,\be,\mu]= Z^\lam[V,\be,\mu]$, another way to show $P'=P$ is to show that the grand potential $\Omega^\lam[V,\be,\mu]$ of $Z^\lam[V,\be,\mu]$ is larger by a factor of $\lambda^D$ than $\Omega[V,\be,\mu]$. Then $\Omega^\lam[V,\be,\mu]=\Omega[\lam^D V,\be,\mu]=\lam^D \Omega[V,\be,\mu] $.\\

%or $-P' \lam^D V=\lam^D (-PV) $ which implies $P'=P$, where $P'$ is the pressure corresponding to $\Omega[\lam^D V,\be,\mu]$. \\

The grand potential $\Omega^\lam$ is given by:

\en
\Omega^\lam=-\beta^{-1} \ln Z^\lam[V,\be,\mu].
\een

By the cluster expansion, $\Omega^\lam$ is given by the sum of connected vacuum graphs. $Z^\lam[V,\be,\mu]$ and $Z[V,\be,\mu]$ have the same macroscopic parameters and only differ in that $Z^\lam$'s propagator is 

\en
\Delta^\lam=\f{1}{i\omega_n-\f{k^2}{2\lambda^2}-\mu},
\een

and that the potential is 

\en
V^\lam \Q(\vec{x}\,-\vec{y}\W)=V(\lambda \Q(\vec{x}\,-\vec{y}\W)\,)
\een

instead of $V\Q(\vec{x}\,-\vec{y}\W)$. Fourier transforming Eq. (14) gives the relationship:

\en
V^\lam \Q(\vec{k}\W)=\f{V\Q( \f{\vec{k}}{\lam}\W)}{\lam^D}
\een

The Feynman rules for the theory say that each vertex contributes its Fourier transform $V^\lam \Q(\vec{k}\W)$, where $\vec{k}$ is the momentum flowing through the vertex, and each propagator contributes Eq. (13). For vacuum graphs, all momenta $\vec{k}$ in the vertices and propagators are integrated over in loop momenta $\int \f{d^Dk}{(2\pi)^D}$. Let us make the change of variables $\int \f{d^Dk}{(2\pi)^D}=\int \lam^D \f{d^Dk}{(2\pi)^D \lam^D}=\int \lam^D \f{d^D\tilde{k}}{(2\pi)^D }$ and relabel $\tilde{k}$ as $\vec{k}$. This will cause $\Delta_\lam(i\omega, \vec{k})=\Delta\Q(i\omega, \f{\vec{k}}{\lam}\W) \rightarrow \Delta(i\omega, \vec{k})$ and $V^\lam\Q(\vec{k}\W)=\f{V\Q( \f{\vec{k}}{\lam}\W)}{\lam^D} \rightarrow \f{V\Q(\vec{k}\W)}{\lam^D} $ in the loop integrals. \\

Therefore, $\Omega^\lam$ is the same as $\Omega$, except for an overall scale factor of $\Q(\f{1}{\lam^D}\W)^\nu \Q(\lam^D\W)^L$, where $\nu$ is the number of vertices and $L$ is the number of loops. Topologically, for connected vacuum graphs of the 2-body potential, $L=\nu+1$. So the overall scale factor becomes $\lambda^D$. Hence $\Omega^\lam=\lam^D \Omega$, and therefore $P'=P$. \\

This generalizes to translationally-invariant n-body potentials, and for spontaneous symmetry breaking. Suppose the interaction is of the form:

\en
\int_{V'} \Q(\prod\limits_{i=1}^n d^D\vec{x}_i' \,\phi'^{m(i)}(\tau,\vec{x}'_i)\W) V(\vec{x}_1',...\vec{x}_n')=\f{\lam^{Dn}}{\lam^\f{DM}{2}}\int_{V} \Q(\prod\limits_{i=1}^n d^D\vec{x}_i \,\phi^{m(i)}(\tau,\vec{x}_i)\W) V(\lam \vec{x}_1,...\lam \vec{x}_n)
\een

where $m(i)$ is the number of fields in the interaction with spatial coordinate $\vec{x}_i$, and $M=\sum\limits_{i=1}^n m(i)$. For translationally-invariant potentials

\en
V^\lam=\f{V\Q(\f{k}{\lam}\W)}{\lam^{D(n-1)}}.
\een

So

\en
\Omega^\lam=\Q(           \f{\lam^{Dn}}{\lam^{\f{DM}{2}}}          \f{1}{\lam^{D(n-1)}}        \W)^\nu \Q(\lam^D\W)^{L}\Omega.
\een

Since $L=\Q(\f{M}{2}-1\W)\nu+1$,\footnote{$M$ lines come out of each vertex, and each line coming out is $1/2$ of an internal line, so $\f{M\nu}{2}=I$ where $I$ is the number of internal lines. The number of loops is the number of independent momenta, $L=I-\nu+1$. So $L=\Q(\f{M}{2}-1\W)\nu+1$.} this again gives:

\en
\Omega^\lam=\lambda^D \Omega .
\een

For a diagram with a mixture of vertices of different types, $L=\sum\limits_{i}\Q(\f{M_i}{2}-1\W)\nu_i+1$, where $\nu_i$ is the number of vertices of type $i$, and $M_i$ is the number of lines coming out of each vertex:

\en
\Omega^\lam&=\Q[\prod\limits_i \Q(           \f{\lam^{Dn_i}}{\lam^{\f{DM_i}{2}}}          \f{1}{\lam^{D(n_i-1)}}        \W)^{\nu_i}\W] \Q(\lam^D\W)^{\sum\limits_{i}\Q(\f{M_i}{2}-1\W)\nu_i+1}\Omega \\
&=\lam^D \Omega.
\een

\begin{comment}
\en
V^\lam=\f{V}{\lam^{(n-1)D}}
\een

leading to an overall scale factor of $\Q(\f{1}{\lam^{(n-1)D}}\W)^V \Q(\lam^D\W)^L$. Topologically, for connected vacuum graphs of the n-body potential, $L=(n-1)V+1$. Hence $\Omega^\lam=\lam^D \Omega$, and therefore $P'=P$. \\

For spontaneously broken theories, you can have $2n+1$ terms coming out of a vertex. The potential term then has an additional factor of $\f{1}{\lambda^{D/2}}$ over the $2n$ term. But the number of loops increases (for a constant number of vertices $V$) by $\f{V}{2}$. So the overall scale factor changes by $\Q(\f{1}{\lam^{D/2}}\W)^V\Q(\lam^D\W)^{\f{V}{2}}=1$
\end{comment}
%Suppose $V(\vec{x}-\vec{y}\,)=V(|\vec{x}-\vec{y}\,|)\equiv V(r)$, and that $V(r)=c_n r^n$ can be expanded via Laurent series with coefficients $c_n$. \\

%Since $Z=e^{\beta PV}$ and $Z^\lam[V,\be^{-1},\mu]=Z[\lambda^D V,\be^{-1},\mu]$, it also follows that $P^\lambda V=P \lambda^D V$ so that $P^\lambda=\lambda^D P$: we will %show this explicitly via diagrams in the appendix.  

\section{Scale Equation}

The virial equation, Eq. (9), can be recast into a different form that illustrates the effect of microscopic scales on the thermodynamics of a system. A simple way to see this is to write the potential as\footnote{We are now restricting ourselves to radial potentials.}:

\en
V(|\vec{x}-\vec{y}\,|)=\f{f\left(\f{g_i}{|\vec{x}-\vec{y}\,|^{[g_i]}}\right)}{|\vec{x}-\vec{y}\,|^2}.
\een

$f$ is a dimensionless function whose arguments are the ratios of the couplings $g_i$ of $V$ to their length dimension $[g_i]$ expressed in units of $|\vec{x}-\vec{y}\,|$ ($\f{\hbar^2}{m}\f{1}{|\vec{x}-\vec{y}\,|^2}$ provides units of energy)\footnote{As an example, consider $V(|\vec{x}-\vec{y}\,|)=\f{k}{2}|\vec{x}-\vec{y}\,|^2+\lambda|\vec{x}-\vec{y}\,|$, where the coupling $k$ has length dimension -4 and $\lambda$ has length dimension -3. Then $f\left(\f{k}{|\vec{x}-\vec{y}\,|^{[k]}},\f{\lambda}{|\vec{x}-\vec{y}\,|^{[\lambda]}}\right)=\f{1}{2}\f{k}{|\vec{x}-\vec{y}\,|^{-4}}+\f{\lambda}{|\vec{x}-\vec{y}\,|^{-3}}$. The couplings $k$ and $\lambda$ provide the characteristic length scales.}. Denoting $r=|\vec{x}-\vec{y}\,|$

\en
r\f{dV}{dr}&=-2V(r)+\f{1}{r}\f{df\left(\f{g_i}{r^{[g_i]}}\right)}{dr} \\
&=-2V(r)-\f{1}{r^2}  [g_i]g_i\f{\p  f\left(\f{g_i}{r^{[g_i]}}\right)}{\p g_i}\\
&=-2V(r)-[g_i] g_i \f{\p V}{\p g_i}.
\een

where the chain rule was used in line 2. Substituting this into Eq. (9) gives

\en
DPV&=2KE+2V-\Q\la \f{1}{2}\int d^D \vec{x}\,\int d^D \vec{y}\, \rho(\tau,\vec{y}\,) \Q( -[g_i]g_i\f{\p V}{\p g_i} \W) \rho(\tau,\vec{x}\,) \W\ra \\
&=2E+\Q\la \f{1}{2}\int d^D \vec{x}\,\int d^D \vec{y}\,\rho(\tau,\vec{y}\,) \Q( [g_i]g_i\f{\p V}{\p g_i} \W)\rho(\tau,\vec{x}\,) \W\ra. 
\een

Rearranging:

\en
2E-DPV=-\Q\la \f{1}{2}\int d^D \vec{x}\,\int d^D \vec{y}\,\rho(\tau,\vec{y}\,)  \Q( [g_i]g_i\f{\p V}{\p g_i} \W)\rho(\tau,\vec{x}\,)  \W\ra. 
\een

On the LHS of Eq. (24) are macroscopic thermodynamic variables. The RHS is a measure of the microscopic physics of the system. In particular, if the potential has no scales $[g_i]=0$ and no anomalies (i.e., $J=1$), you get 0 on the RHS, and Eq. (24) reduces to the equation of state for a non-relativistic scale-invariant system \cite{italians}.

\begin{comment}in a Laurent expansion with coefficients $c_n$. Then:

\en
r\f{dV}{dr}=n c_n r^{n}
\een

Noting that the length dimension ($\hbar=m=1$) of $c_n$ is $[c_n]=-n-2$. Therefore Eq. (14) can be re-written as 

\en
r\f{dV}{dr}&=-\Q([c_n]+2\W) c_n r^{n} \\
&=-[c_n]\f{\p V}{\p c_n}-2V
\een
 \end{comment}

\section{Conclusion and Comments}
The goal of this paper has been to highlight certain features in the derivation of the virial theorem for non-relativistic systems, which display a potentially important omission due to the presence of the Jacobian needed in the path-integral derivation developed here.  Indeed, while we set $J=1$ at the outset in order to make contact with the literature (specifically, Toyoda's et al. work \cite{t1,t2,t3}), Eq. (6) shows that the natural procedure would be to not assume this and keep the contribution of the Jacobian, regardless of whether or not there is a classical scaling symmetry. Obviously, in the latter case, one has to keep the Jacobian in order to incorporate the quantum anomaly as was shown in \cite{Ord}. The formal mathematical steps in the general case presented here are the same as in that paper, and Eq. (24) would become

\en
2E-DPV=-\Q\la \f{1}{2}\int d^D \vec{x}\,\int d^D \vec{y}\,\rho(\tau,\vec{y}\,)  \Q( [g_i]g_i\f{\p V}{\p g_i} \W)\rho(\tau,\vec{x}\,)  \W\ra -\f{1}{\beta}\hat{\text{T}}\text{r} \Q(\hat{\theta}_s \de(\tau_x-\tau_y) \delta^{D}(\vec{x}-\vec{y}\,)I_2 \W),
\een

where $I_2=\pma 1 & 0 \\ 0 & 1\epma$, $\hat{\theta}_s =-\Q(1+\vec{x}\cdot \vec{\nabla} \W)$, and we have also used the $2\times 2$ matrix notation of \cite{Lin} ($\hat{\text{T}}\text{r}$ includes both a matrix and functional trace). \\

%$\hat{\text{T}}\text{r}$ is a functional trace over 

As with the work in \cite{Ord} and \cite{Lin}, the key to assess the importance of the Jacobian term rests upon one's ability to compute its contribution in detail, which implies a careful regularization procedure, and possibly also renormalization. The actual details will depend of the type of potentials considered.  An interesting direction is the relativistic generalization of these ideas. Work on this is currently in progress \cite{prog}. 

\begin{comment}
From Eq. (5), we can see that Jacobian comes naturally out of the path integral formalism. So the anomalous virial theorem comes naturally, and we had to set $J=1$, which is unnatural, to recover the non-anomalous virial theorem. \\

Work in progress includes the relativistic case, and further analysis of the structure of Eq. (17).
\end{comment}

\begin{acknowledgments}
One of us (CRO) wishes to thank the Technological University of Panama for its hospitality at different stages of this work.
\end{acknowledgments}

%\bibliography{vir}
\end{document}